\documentclass[12pt,preprint]{aastex}

\def\alwaysmath#1{\ifmmode{#1}\else{$#1$}\fi} 
 
\lefthead{Majewski et al.} 
\righthead{APOGEE Observations of the Sagittarius dSph Galaxy}

\begin{document}

\title{Discovery of a Dynamical Cold Point\\
 in the Heart of the Sagittarius dSph Galaxy\\
with Observations from the APOGEE Project} 
 
\author{Steven R. Majewski\altaffilmark{1},
Sten Hasselquist\altaffilmark{1,2},
Ewa L. {\L}okas\altaffilmark{3}, 
David L. Nidever\altaffilmark{1,4}, \\
Peter M. Frinchaboy\altaffilmark{5},
Ana E. Garc\'{\i}a P\'erez\altaffilmark{1},
Kathryn V. Johnston\altaffilmark{6},
Szabolcs M{\'e}sz{\'a}ros\altaffilmark{7},\\
Matthew Shetrone\altaffilmark{8}, 
Carlos Allende Prieto\altaffilmark{7},
Rachael L. Beaton\altaffilmark{1},
Timothy C. Beers\altaffilmark{9,10},\\
Dmitry Bizyaev\altaffilmark{11},
Katia Cunha\altaffilmark{9,12,13},
Guillermo Damke\altaffilmark{1},
Garrett Ebelke\altaffilmark{11},\\
Daniel J. Eisenstein\altaffilmark{14},
Fred Hearty\altaffilmark{1},
Jon Holtzman\altaffilmark{2},
Jennifer A. Johnson\altaffilmark{15},\\
David R. Law\altaffilmark{16},
Viktor Malanushenko\altaffilmark{11},
Elena  Malanushenko\altaffilmark{11},
Robert W. O'Connell\altaffilmark{1},\\
Daniel Oravetz\altaffilmark{11},
Kaike Pan\altaffilmark{11},
Ricardo P. Schiavon\altaffilmark{17,18},
Donald P. Schneider\altaffilmark{19},\\
Audrey Simmons\altaffilmark{11},
Michael F. Skrutskie\altaffilmark{1},
Verne V. Smith\altaffilmark{9},
John C. Wilson\altaffilmark{1},\\
Gail Zasowski\altaffilmark{1,15}}
 
\altaffiltext{1}{Dept. of Astronomy, University of Virginia, 
Charlottesville, VA 22904-4325, USA (srm4n, sh6cy, aeg4x, rlb9n, gjd3r, frh3z, rwo, jcw6z@virginia.edu)}

\altaffiltext{2}{New Mexico State University, Las Cruces, NM 88003, USA (holtz@nmsu.edu)}

\altaffiltext{3}{Nicolaus Copernicus Astronomical Center, Bartycka 18, 00-716 Warsaw, Poland (lokas@camk.edu.pl)}

\altaffiltext{4}{Department of Astronomy, University of Michigan, Ann Arbor, MI 48109, USA (dnidever@umich.edu)}

\altaffiltext{5}{Texas Christian University, Fort Worth, TX 76129, USA (p.frinchaboy@tcu.edu)}

\altaffiltext{6}{Department of Astronomy, Columbia University, Mail Code 5246, New York, NY 10027, USA (kvj@astro.columbia.edu)}

\altaffiltext{7}{Instituto de Astrof\'{\i}sica de Canarias, 38205 La Laguna, Tenerife, Spain;
Departamento de Astrof\'{\i}sica, Universidad de La Laguna,
38206 La Laguna, Tenerife, Spain (meszi, callende@iac.es)}

\altaffiltext{8}{University of Texas at Austin, McDonald Observatory, Fort Davis, TX 79734, USA
(shetrone@astro.as.utexas.edu)}

\altaffiltext{9}{National Optical Astronomy Observatories, Tucson, AZ 85719, USA (cunha, vsmith@email.noao.edu)}

\altaffiltext{10}{Department of Physics \& Astronomy and JINA, Joint Institute for Nuclear Astrophysics, Michigan State University, E. Lansing, MI  48824, USA (beers@pa.msu.edu)}

\altaffiltext{11}{Apache Point Observatory, P.O. Box 59, Sunspot, NM 88349-0059, USA (dmbiz, gebelke, viktorm, elenam, doravetz, kpan, asimmons@apo.nmsu.edu)}

\altaffiltext{12}{Steward Observatory, University of Arizona, Tucson, AZ 85721, USA} 

\altaffiltext{13}{Observat\'orio Nacional, S\~ao Crist\'ov\~ao, Rio de Janeiro, Brazil}

\altaffiltext{14}{Harvard-Smithsonian Center for Astrophysics, 60 Garden St., MS \#20,
Cambridge, MA 02138, USA (deisenstein@cfa.harvard.edu)}

\altaffiltext{15}{Department of Astronomy, The Ohio State University, Columbus, OH 43210, USA ( jaj@astronomy.ohio-state.edu, gail.zasowski@gmail.com)}

\altaffiltext{16}{Dunlap Institute for Astronomy and Astrophysics, University of Toronto, 50 St. George Street, Toronto, Ontario M5S 3H4, Canada (drlaw@di.utoronto.ca)}

\altaffiltext{17}{Gemini Observatory, 670 N. A'Ohoku Place, Hilo, HI 96720, USA (rpschiavon@gmail.com)}

\altaffiltext{18}{Astrophysics Research Institute, Liverpool John Moores University, Twelve Quays House, Egerton Wharf, Birkenhead CH41 ILD, United Kingdom}

\altaffiltext{19}{Department of Astronomy and Astrophysics, The Pennsylvania State University,
  University Park, PA 16802; Institute for Gravitation and the Cosmos, The Pennsylvania State University, University Park, PA 16802, USA (dps7@psu.edu)}

\begin{abstract} 
 
The dynamics 
of the core of the Sagittarius (Sgr) dwarf spheroidal (dSph) galaxy
are explored using high-resolution ($R$$\sim$$22,500$), $H$-band, near-infrared spectra of
over 1,000 giant stars in the central 3 deg$^2$ of the system, of which 328 are identified as Sgr members.
These data, 
among some of the earliest observations from the SDSS-III/Apache Point Observatory Galactic Evolution
Experiment (APOGEE) and the largest published sample of high resolution Sgr dSph spectra 
to date, reveal 
a distinct gradient
in the velocity dispersion of Sgr from 11-14 km s$^{-1}$ for radii $>0.8^{\circ}$ from center
to a dynamical cold point of 8 km s$^{-1}$ in the Sgr center 
--- a trend differing from that found in previous kinematical analyses of Sgr 
over larger scales
that suggest a more or less flat dispersion profile at these radii.  Well-fitting mass  
models with either cored and cusped dark matter distributions can be found to match the
kinematical results, although the cored profile succeeds with significantly more isotropic stellar orbits than
required for a cusped profile.  It is unlikely that the cold point reflects an unusual
mass distribution. 
The dispersion gradient may arise from
variations in the mixture of populations with distinct kinematics within the
dSph; this explanation is suggested (e.g., by detection of a metallicity gradient across
similar radii), but not confirmed, 
by the present data.
Despite these remaining uncertainties about their interpretation,
 these early test data (including some from instrument commissioning) 
 demonstrate APOGEE's usefulness
 for precision dynamical studies, even 
for fields observed at extreme airmasses.
 
\end{abstract}

 
\section{Introduction} 
 
The Sagittarius (Sgr) dSph galaxy is a compelling Milky Way (MW) satellite for intense study,
given (1) its unusual star formation and chemical enrichment history 
(e.g., Smecker-Hane
\& McWilliam 2002, 
Siegel et al. 2007, 
Chou et al. 2010)
and other properties (e.g., its own 
globular cluster system)
evoking similarities to the Magellanic Clouds (e.g., Chou et al. 2010, {\L}okas et al. 2010a), but also (2) some
remarkably unique physical properties.  For example, Sgr presents the most vivid example of the MW's
hierarchical growth via satellite accretion.  It is also the only clear example of a nucleated dwarf
galaxy among MW satellites, with a prominent metal-poor globular cluster (M54) possibly coinciding
in phase space with its nucleus  (e.g., Layden \& Sarajedini 2000, Majewski et al. 2003, Monaco
et al. 2005a, Bellazzini et al. 2008 [B08]; but cf. Siegel et al. 2011).\footnote{That $\omega$ Centauri represents the remains of a similar,
nucleated dwarf galaxy with a superposed globular cluster is an intriguing,
but as yet not fully proven, hypothesis (B08, Carretta et al. 2010).}

The latter fact is particularly germane to the debate over the dark matter distribution
in dSph systems: While prevailing cold dark matter (CDM) models predict that their total density 
profiles should be cusped, their luminous density profiles are cored.  Furthermore, several dynamical 
assessments of dSphs 
favor cored mass profiles: 
\begin{enumerate}
\item Kleyna et al. (2003) argue that the double-peaked stellar structure of the
Ursa Minor dSph (e.g., 
Palma et al. 2003)
could only have 
survived a Hubble time if it lived within a host possessing a cored mass profile, 
whereas the dynamically cold, secondary clump would
have been erased in less than a Gyr within a cusped mass distribution.
\item Calculations suggest that if the Fornax dSph had a cuspy mass profile its 
five globular clusters would have sunk
to the center by dynamical friction in much less than a Hubble time (Goerdt et al. 2006, 
S\'anchez-Salcedo et al. 2006).
\item Analysis of combined surface brightness profiles and velocity dispersions for some dSph galaxies
lead to a preference for lower density, cored mass distributions (e.g., Gilmore et al. 2007,
Battaglia et al. 2008).
\end{enumerate}
If these implied cored profiles and inferred
low central phase space densities are primordial and not the result of  
modification during the dynamical life of the satellite, a warm dark matter particle is implied
(e.g., gravitinos, light sterile neutrinos), rather than one of the WIMP candidates of CDM
(e.g., axions, neutralinos).  On the other hand, B08 use N-body simulations 
to argue that the presence of M54 
in the very center of Sgr is compelling evidence for a cusped profile, which would
have spiraled M54 to the nucleus by dynamical friction within a Hubble time
if M54 had been born anywhere within $\sim$5 kpc of center (see also 
Monaco et al. 2005a).  However, isochrone fitting to precision photometry
of M54 and the Sgr core from ACS on the Hubble Space Telescope
yields a 2 kpc distance difference, implying that M54 may not be {\it at} the 
Sgr core, but merely projected upon it (Siegel et al. 2011).
Clearly further work is needed to elucidate the true density profile shape for Sgr 
(and, by analogy, 
other dSphs).
Extensive, high accuracy velocity mapping of the dSph is expected to provide
further data needed to help discriminate between competing mass models.

Nevertheless, despite Sgr's proximity, 
this dSph has seen surprisingly scant attention in terms of 
high resolution spectroscopic study compared to other, more distant MW satellites.  
Because the system is of large angular size, even with the $\sim$$1$ deg$^2$ 
fields-of-view (FOV) typical of many multifiber spectrographs, only pencil-beam samplings
are possible, 
and these are typically at lower resolution.  The most comprehensive Sgr spectroscopic
studies at any resolution are those of 
Ibata et al. (1997, $R$$\sim$$5000$), 
B08 ($R$$\sim$$5500$),
Penarrubia et al. (2011,  $R$$\sim$$10,000$), and
Frinchaboy et al. (2012,  $R$$\sim$$15,000$ --- ``F12" hereafter),
which together probe 24 independent
directions and $\sim$3700 different stars.  
The largest survey, F12,
covers only $\sim$$10$\% of the area within the isopleth corresponding to Sgr's 1800 arcmin
King limiting radius (Majewski et al. 2003).
Meanwhile, only a few dozen total stars have been explored at ``echelle" resolution
across the dwarf by the combined studies of Smecker-Hane
\& McWilliam (2002), Monaco et al. (2005b), 
Chou et al. (2010) and Sbordone et al. (2007).\footnote{While
B08's Sgr+M54 survey did include some stars observed at $R$$\sim$$17,000$ with FLAMES, these were 
primarily M54 stars and their entire survey was concentrated within 9' of center.}
As a result, we know comparably little about the detailed abundance distributions and dynamics of the
core of this intriguing system compared to those of other classical MW
dSphs where velocity dispersion profiles and chemical abundance patterns are derived with
hundreds to thousands of stellar members (e.g., Walker et al. 2007)


This situation will surely soon change if any of the 
several 
available multiobject, high resolution spectrographs are devoted to large-scale surveys of this most interesting Galactic satellite.
In the meantime, a significant high resolution spectroscopic sample of spectra of Sgr 
has already been obtained
by the Sloan Digital Sky Survey III (SDSS3; Eisenstein et al. 2011) 
as a byproduct of commissioning and early survey observations by 
the Apache Point Observatory
Galactic Evolution Experiment (APOGEE).  These high resolution, $H$-band spectra 
lie at interesting positions probing Sgr
between the intense central nucleus study of B08
and the larger radii probed by most other
Sgr surveys.  We use these APOGEE spectra to show, for the first time, that within
$1^{\circ}$ of its center the Sgr dSph contains 
a strong velocity dispersion gradient and a modest [Fe/H] gradient.

\section{APOGEE Spectra of Sgr and Their Reduction}

The APOGEE project is described in Allende Prieto et al. (2008), Majewski et al. (2010), 
and Eisenstein et al. (2011).
It uses a cryogenic, bench-mounted spectrograph recording 300 simultaneous
spectra covering $\lambda\lambda$ 1.51-1.70$\mu$m for light fed to it
from the Sloan 2.5-m telescope (Gunn et al. 2006) via 40-m long optical fibers.
The spectrograph, described by Wilson et al. (2010),
records three distinct spectral regions onto separate detectors (spanning 1.51-1.58, 1.59-1.64
and 1.65-1.70 $\mu$m respectively)
at§ $R\sim22,500$. 

\begin{figure}[th]
\epsscale{1.}
\plotone{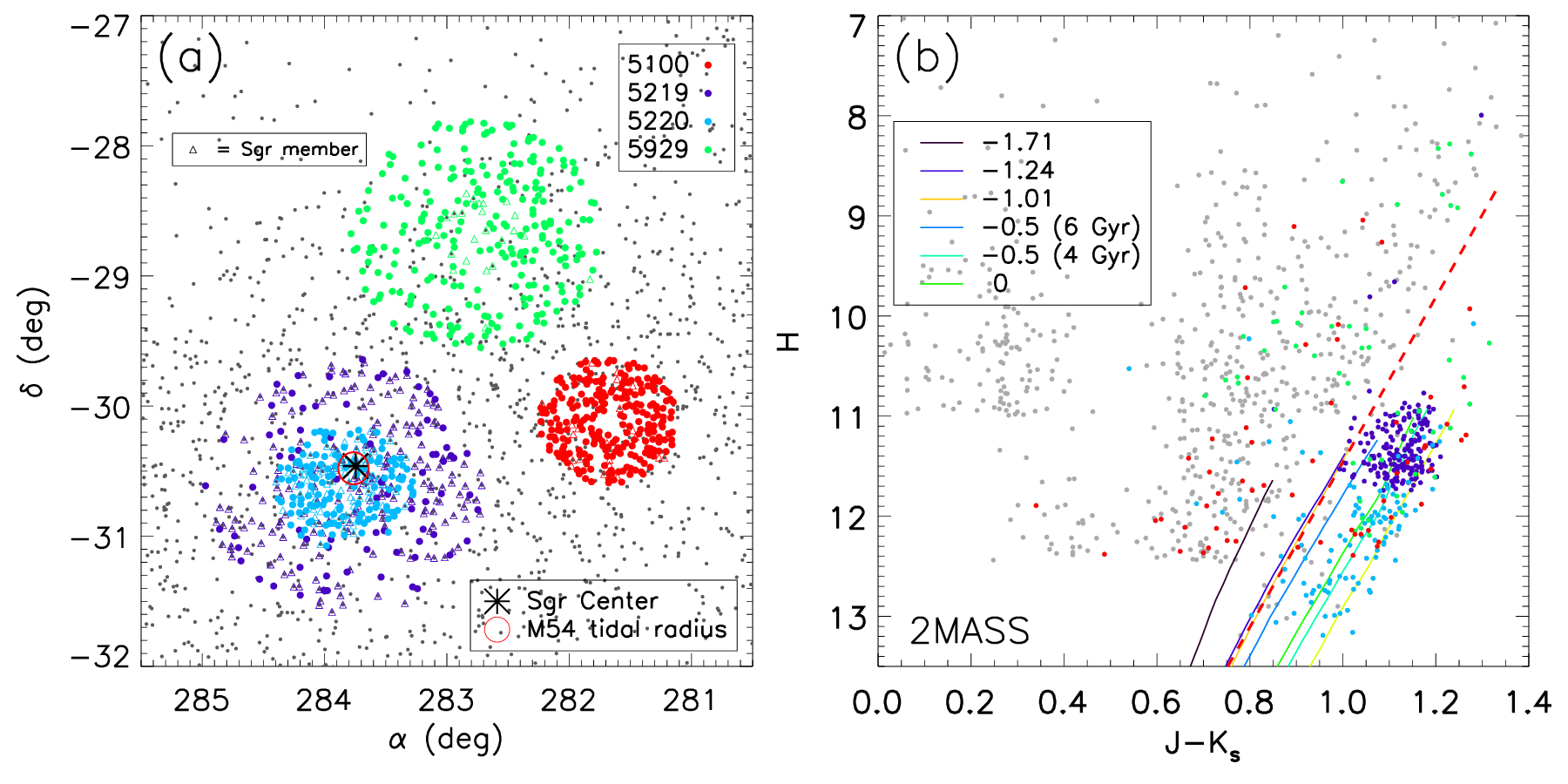} 
\caption{The sky ({\it left}) and 2MASS color-magnitude ({\it right}) distributions of observed
stars in Sgr, color-coded by 
plugplate number (but only for stars with $90<v_{GSR}<220$ km s$^{-1}$
 in the right panel).  The left panel also shows the Majewski et al. (2003) distribution of 
 M giant stars ({\it grey points}) and fitted center ({\it black star}) for Sgr as well as 
 the 7.4' tidal radius of M54 (B08).
  The CMD 
includes Padova isochrones (Bressan et al. 2012) corresponding to major Sgr populations
identified by
Siegel et al. (2007) and our limit for accepting potential Sgr members ({\it red dashed line}). }
\end{figure}

The 2 arcsec diameter, APOGEE fibers 
are plugged into standard Sloan plugplates and
observed similarly to the optical Sloan surveys (e.g., Smee et al. 2013), 
but with these
variations (see Zasowski et al. 2013): 
(1) 35 fibers in each plugplate configuration are used to collect sky/airglow spectra, 
(2) another 35 fibers target bright ($5 < H < 11$), 
hot (generally $[J-K_s] < 0.4$) stars to monitor telluric (H$_2$O, CO$_2$, CH$_4$)
absorption, and 
(3) because the extreme zenith distances at APO required to observe the Sgr core 
($63^{\circ}$) 
 impose strong differential refraction effects, Sgr
plugplates were tested with various 
field sizes and magnitude limits.  
Table 1 (see also Fig.~1) summarizes the Sgr core plugplates observed and when, 
their field centers and diameters (FOV), 
$H$-band limit ($H_{lim}$) for stars targeted with standard APOGEE criteria (see below)
and for the faintest Sgr member, 
total integration time,
number of Sgr members ending up in our final dynamical analysis (see below),
total number of stars with measured radial velocities (RVs), 
and their median RV error.
Observations of plate 5100
were taken before the instrument was in its fully commissioned state and with
elements in non-optimal alignment, which led  to
slightly blurred line-spread-functions degrading the resolution 
of the 1.65-1.70 $\mu$m region to only $R\sim14,500$;
therefore, to use the highest quality spectral regions and maintain consistency, 
analysis of all spectra was
confined to 1.51-1.64 $\mu$m, a region where, in any case, 
relevant information density happens to concentrate.
Targets in plates 5219, 5220, and 5929 were selected from Sgr members identified by
F12, with leftover fibers filled by random sampling
$(J-K)_0 \ge 0.5$ stars from the 2MASS point source 
catalog\footnote{The 2MASS photometry was dereddened using the RJCE method 
(Majewski et al. 2011) with WISE (Wright et al. 2010) providing the required
mid-infrared photometry.} as described in Zasowski et al. (2013);
targets in plate 5100 were selected using only the latter method (Fig.~1).
 
\begin{table}[htdp]
\caption{APOGEE Observations in the Field of the Sagittarius dSph Galaxy}
\begin{center}
\small
\begin{tabular}{clrccccccc}
\hline
plate & ~~UT date & ($l$,$b$) (deg) & FOV & $H_{lim}$(plate/Sgr)  &  int(s) & members & RVs & $\epsilon_{RV}$ (km s$^{-1}$)\\
\hline
5100	 & 2011-7-18   & (5.2,-12.3)	&  1$^{\circ}$   & 12.5/12.4 & 4729	&   ~24 & 262 & 0.14\\
5219 & 2011-9-6,7 & (5.5,-14.2)	& 2$^{\circ}$    & 11.0/11.7 & 8004	&  190  & 265 & 0.15\\
5220 & 2012-6-5,12 & (5.5,-14.2)	& 1$^{\circ}$    & 12.2/13.2 & 8004	&   ~91 & 218 & 0.30\\
5929	 & 2012-6-6 	   &(6.9,-12.6)	& 2$^{\circ}$    & 11.0/12.2 & 4002	&   ~23 & 262 & 0.09\\
\hline
\end{tabular}
\end{center}
\label{default}
\end{table}%


An automated data reduction pipeline written specifically for APOGEE (Nidever et al., in 
preparation) was used to convert the raw APOGEE datacubes into 1-D, wavelength-calibrated
spectra and derive RVs previously shown to be accurate to $0.26\pm0.22$
km s$^{-1}$ (see Nidever et al. 2012).
For 46 stars in common between plugplates 5219 and 5220 the
dispersion in difference between derived RVs is 0.29 km s$^{-1}$.

Metallicities ([Fe/H]) are derived using an automated method with the APOGEE
Stellar Parameters and Chemical Abundances Pipeline (ASPCAP;
Garc\'{\i}a P\'erez et al., in preparation). The version of 
ASPCAP code used for the present analysis 
derives stellar atmospheric parameters via $\chi^2$ fits of the airglow-masked 
APOGEE spectra against spectra interpolated in 
libraries 
of synthetic spectra at the
observed resolution.  For the K and M spectral classes of interest here the code uses a six-dimensional
library with dimensions spanning ranges: 
(1) $3,500$$\le$$T_{\rm eff}$$\le$$5,000$ K, 
(2) $0.0$$\le$$\log{g}$$\le$$5.0$, 
(3) $-2.5$$\le$[Fe/H]$\le$$+0.5$, 
(4) $-1.0$$\le$[C/Fe]$\le$$+1.0$, 
(5) $-1.0$$\le$[N/Fe]$\le$$+1.0$, and 
(6) $-1.0$$\le$[$\alpha$/Fe]$\le$$+1.0$. 
The library is based on ATLAS9 model 
atmospheres (Castelli \& Kurucz 2004) and spectral synthesis calculations with the code 
ASS$\epsilon$T (Koesterke et al. 2008)
and an atomic line list optimized to match the solar spectrum.
The microturbulence was fixed at $\xi=2$ km s$^{-1}$.
Tests of the [Fe/H] delivered by this version of ASPCAP code were made via
APOGEE observations of 20 open and globular clusters having high quality metallicities
in the literature.  From this assessment we find that the
ASPCAP [Fe/H] are reliable to 0.06-0.10 dex when compared to literature values at solar 
metallicity, and to 0.10-0.15 dex at [Fe/H]$\le$$-1$.

 \begin{figure}[th]
\epsscale{0.9}
\plotone{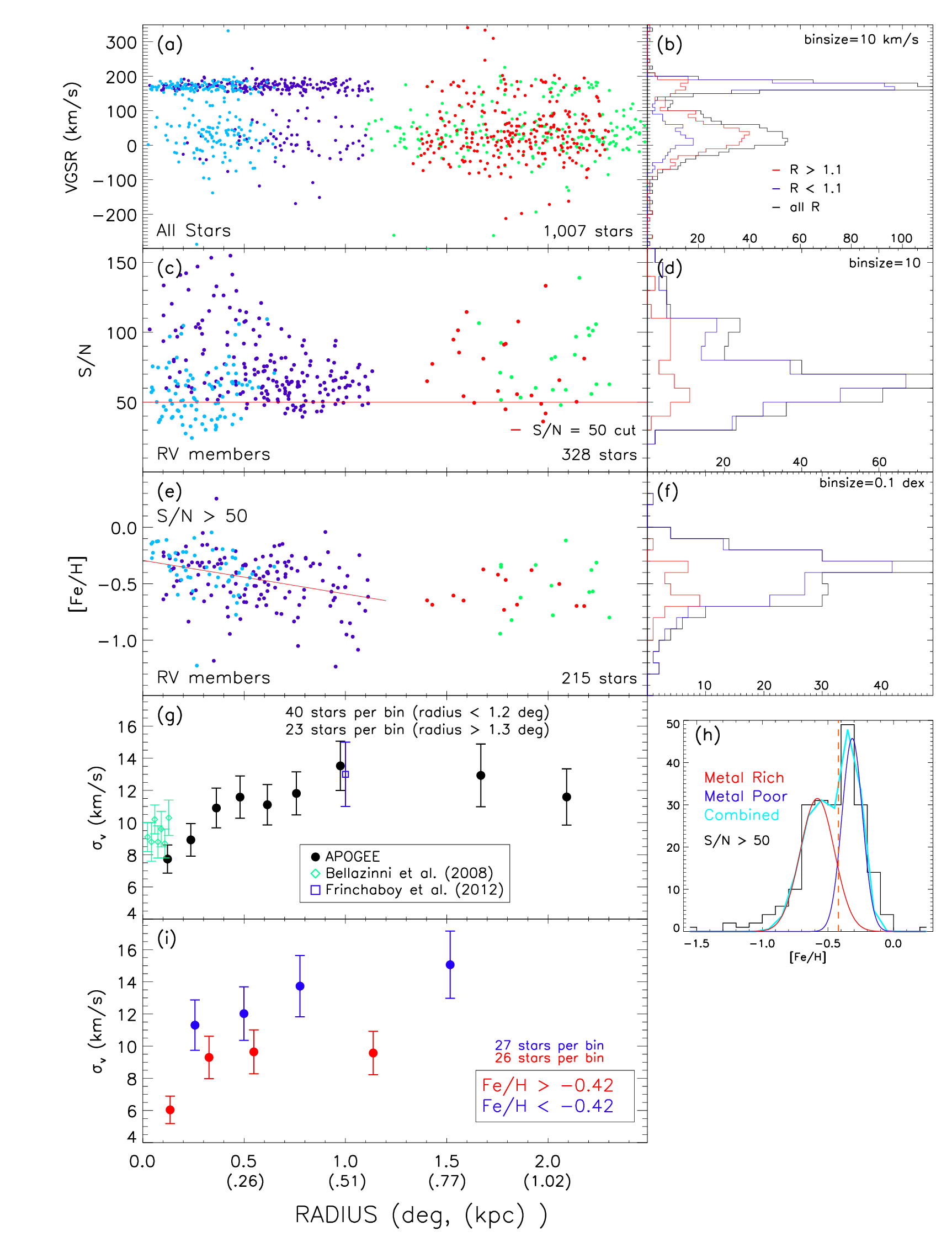}
\caption{Measured parameters from APOGEE spectra of the Sgr core, with
points ({\it left panels}) color-coded as in Fig.~1 and the distributions of those
parameters  ({\it panels b, d and f}) in distinct histograms for the core ({\it blue}) and
offset ({\it red}) fields.  
Panels {\it g} and {\it i} show our derived $\sigma_v$ gradient for the entire sample 
and two metallicity subsamples (defined by the Gaussian fits in panel {\it h}), 
respectively, compared to results from F12 and B08 
({\it panel g}).
}
\end{figure}

\section{Dynamics of Sagittarius Core Stars} 

Figures 2a,b show the distribution of  
Galactic Standard of Rest (GSR) RVs ($v_{GSR}$) for the APOGEE observations of 1,007 stars in 
the Sgr fields, assuming a solar motion in right-handed Galactic coordinates of 
$(+10.0,+225.3,+7.2)$ km s$^{-1}$.  
The different relative strengths of the narrow ``Sgr
peak'' in the inner versus outer two fields (Fig.~2b)
reflects the relative densities of Sgr to MW field stars at the two radii
and the fact that known Sgr stars were deliberately targeted only in the center-field observations.

To determine metallicity and velocity dispersion profiles for Sgr, we  
winnow the data to a more pure sample of Sgr members
by first applying a broad kinematical selection of stars in  
the ``Sgr peak" ($90<v_{GSR}<220$ km s$^{-1}$; colored points in the Fig.~1 CMD) 
followed by
a color-magnitude limit (Fig.~1) roughly tracking an [Fe/H]$=$$-1$ isochrone. 
To this sample we then apply a $3\sigma_v$ iterative outlier rejection scheme to the velocities.
The color and $3\sigma_v$ constraints actually remove
a relatively small fraction of stars in the ``Sgr velocity peak", 
but help reduce contamination from M54 (defined by B08 to span $-1.8$$<$[Fe/H]$<$$-1.1$, and
eliminating one
star within 7.4' of M54 center) and
MW field stars, at the risk of a slight, [Fe/H]$\gtrsim$$-1$ bias (see isochrones in Fig.~1).
All 328 stars remaining have
sufficient $S/N$ (Fig.~2c,d) to have extremely reliable RVs (precision $<<1$ km s$^{-1}$)
for measuring velocity dispersions ($\sigma_v$).

The resulting Sgr $\sigma_v$ profile (Fig.~2g)
shows dramatically, and for the first time, that the heart of Sgr
is characterized by a steady and definitive {\it gradient} from 11-14 km s$^{-1}$ 
for stars at 
$>0.8^{\circ}$ radius to a dynamically colder center at $<8$ km s$^{-1}$ (and with no perceptible rotation).  
As summarized by F12, Sgr velocity dispersions measured by other surveys 
are generally higher than most values shown in Figure 2g, particularly
for at least the next several degrees in radius.
However, only the studies of Ibata et al. (1997) and F12 have sufficient
statistics to derive velocity dispersions in multiple Sgr fields, with
the F12 data of
much better quality due to higher spectral resolution ($R$$=$$15,000$ versus
$5,000$ in Ibata et al.). 
For comparison, we include in Figure 2g the $\sigma_v$ measures of those nearby fields with the 
best data from F12
and B08
(who analyzed 1152 VLT/FLAMES and Keck/DEMOS
spectra of  M54 and the inner 9$\arcmin$ of Sgr).    
The APOGEE observations seemingly ``connect" the warmer F12 and colder B08 points, though we find
a somewhat smaller central Sgr velocity dispersion
than B08, which may relate to APOGEE's order of magnitude smaller intrinsic velocity uncertainties.
The observed gradient should not
be the result of M54 contamination, given our elimination of [Fe/H]$\lesssim$$-1$ stars
by the CMD selection and
M54's tidal radius of $\sim7.4$ arcmin (B08; Fig.~1a).

\section{Explaining the Dynamical Cold Point} 

That Sgr has a strong increase in velocity dispersion with radius makes it 
similar to other MW dSphs with analagous
$\sigma_v$ trends, such as Draco, Sextans and Carina (Walker et al. 2007), although
the causes for such features likely vary.   
For example, Munoz et al. (2008) have suggested that in Carina the phenomenon is related 
to tidal disruption, and while Sgr is also obviously tidally disrupting, the radii that would be affected 
by this should be much farther out than the region probed here (the Fig.~2g gradient 
lies within Sgr's innermost 0.5 kpc, well inside its several kiloparsec
tidal radius; Law \& Majewski 2010).

However, the observed gradient in the velocity dispersion profile of Sgr might naturally 
be explained in terms of
its underlying mass distribution.
Figure 3 shows the profiles of velocity dispersion and a kurtosis-like variable 
($k = [\log(3\kappa/f)]^{1/10}$ where $\kappa$ is the standard estimator of the 
RV distribution kurtosis, and
the correction $f$ depends on the number of stars per bin --- see 
Appendix of {\L}okas \& Mamon 2003)
calculated from a combination of our data and the best data from F12. 
The higher order moment of velocity dispersion represented by $k$
offers an additional sensitive constraint on the mass models.
Together, these data were fit
by Jeans equations solutions (see 
{\L}okas et al. 2005) assuming
a two-component model with the stellar distribution approximated by a S\'{e}rsic profile (with
parameters from {\L}okas et al. 2010a) and the dark matter distribution by
$\rho = C r^{-\alpha} \exp (-r/r_b)$. We considered two inner slopes, $\alpha=0$ (core) and $\alpha=1$ (cusp),
and adjusted the total dark mass, cut-off scale $r_b$ and stellar orbit anisotropy parameter, $\beta$, assumed to be constant with radius.

The cuspy dark matter profile (green lines in Fig.~3) fits the data slightly better than
the profile with the core (red lines in Fig.~3) with $M/L$ within 
$R\lesssim600$ arcmin of 40.2 versus 33.8 
and reduced $\chi^2$ of 1.3 versus 1.7 
(although
neither model seems to match well for $R\gtrsim600$ arcmin).
In spite of resorting to modeling the higher velocity moments, the degeneracy between the parameters
of the dark matter profile (inner slope and characteristic scale) remains (see {\L}okas \& Mamon 2003;
Agnello \& Evans 2012).
However, the best fit of the cored profile is achieved with more isotropic orbits ($\beta= -0.3$)
than for the cuspy one ($\beta = -0.7$). Interestingly, isotropic, or even mildly radial
($\beta \gtrsim 0$) stellar orbits
are predicted in the context of the tidal stirring model for the formation of dSph
galaxies in the Local Group. As shown in {\L}okas et al. (2010b) and Kazantzidis et al. (2011),
strongly tidally-affected dwarfs are dominated by mildly radial orbits as a result of bar formation
at the first pericenter passage on their orbit around the MW. This should also be the case for Sgr
if it formed from a disky progenitor, as proposed by {\L}okas et al. (2010a).

 \begin{figure}[th]
\epsscale{1.0}
\plotone{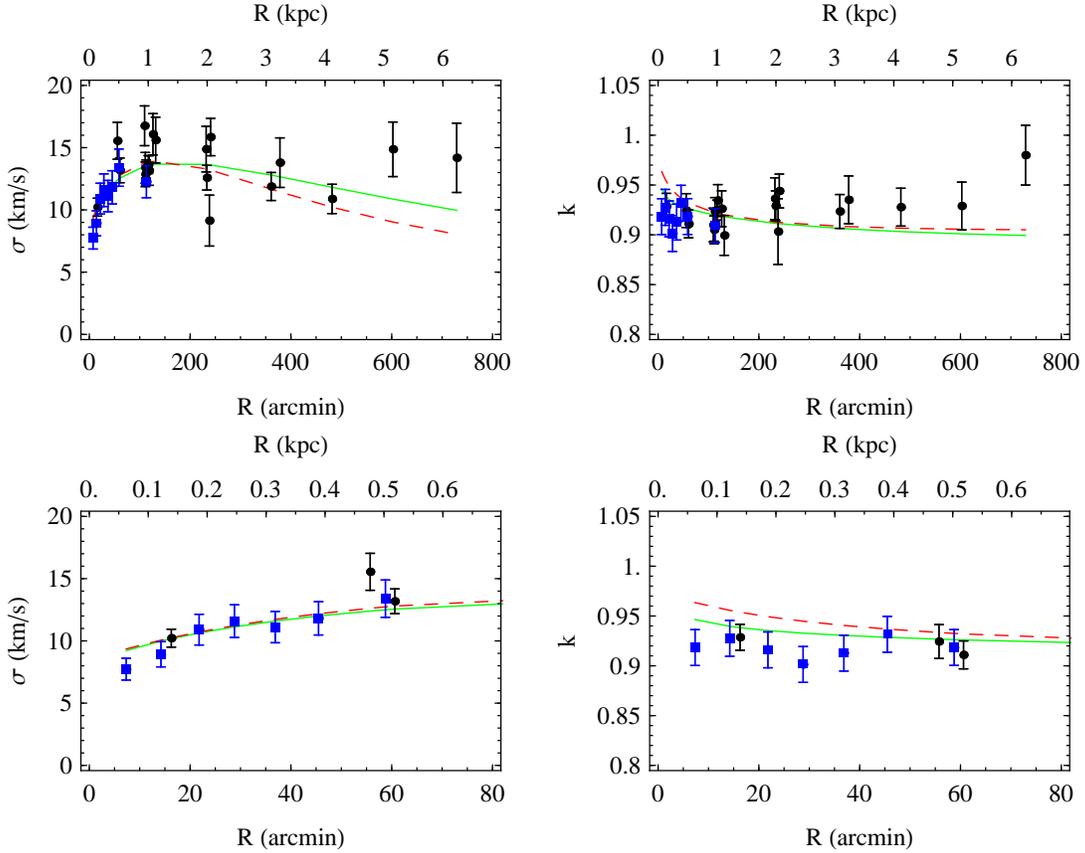}
\caption{Velocity dispersion (left panels) and kurtosis (right panels) data
for the Sgr core from F12
({\it black circles}; we exclude MW-contaminated fields discussed by F12),
and from APOGEE data ({\it
  blue squares}).  The {\it green solid} ({\it red dashed}) line is the best fit of a two-component (stars and dark matter) model
with cuspy (cored) dark matter distribution and constant anisotropy parameter, $\beta$. The best-fitting cuspy (cored)
model has total mass $8.1 (5.6) \times 10^8 M_{\odot}$ and $\beta= -0.7 (-0.3)$.}
\end{figure}

Another explanation suggested (e.g., McConnachie et al. 2007)
to account for systematically smaller velocity dispersions in dSph cores is the
presence of variations in the mixing of  distinct stellar populations with differing
kinematics.  MW dSph galaxies are commonly found to have multiple populations
(e.g., Majewski et al. 1999, Harbeck et al. 2001, 
Tolstoy et al. 2004; 
Ibata et al. 2006),
and Sgr is no exception (e.g., Siegel et al. 2007).  Moreover, from the APOGEE data,
we verify that, at minimum, there is a small but identifiable metallicity gradient within
the inner few degrees of Sgr (Fig.~2e), evident as a few tenths of a dex drop in [Fe/H]
from the center to offset APOGEE fields (Fig.~2f).
To assess this [Fe/H] distribution we have used
only those 215
stars with the most reliable ASPCAP data:
those with $T_{\rm eff}$$>$$3575$ K
and having spectra with total $S/N$$>$$50$ per pixel.
The existence of an [Fe/H] gradient is consistent with 
previous analyses of the Sgr system
(e.g., Alard 2001, Chou et al. 2010)
that show an overall general metallicity 
increase towards the Sgr center.

Although slightly shaped by the combined effects of the color-magnitude selection 
and the ASPCAP temperature limit discussed above, 
the bulk of the
stars in this pruned APOGEE sample have metallicities distributed across
the $-0.8$$<$[Fe/H]$<$$-0.2$ range identified with intermediate age and metallicity stars (``SInt")
and exhibiting several subpopulationZZ 
in the population synthesis of Siegel et al. (2007).
The APOGEE 
metallicity distribution (Fig.~2h) also hints at possible subgroups, and a two-Gaussian fit suggests
[Fe/H]$\sim$$-0.42$ as a reasonable, though approximate, division between them.
These sample subdivisions show some differences in $\sigma_v$ profile,
with the metal-poor subsample exhibiting consistently 
larger $\sigma_v$ within the central few degrees of
Sgr compared to the metal-rich subsample.
That each metallicity subsample still internally shows a dispersion {\it gradient} may belie
the fact that our simple basis for separating population
subsamples is not the cleanest.
On the other hand, the Figure 3 fits find velocity distributions in each radial bin 
consistent (within the errors) with Gaussian (defined as $k=0.93$), whereas
McConnachie et al. (2007) suggest that dwarf galaxies having mixed stellar populations in equilibrium
should yield leptokurtic velocity distributions at radii where multiple populations make a 
significant contribution.

Derivations of the abundances of
additional chemical elements expressed within the APOGEE spectra will hopefully improve
our ability to sort Sgr core stars by population
and test whether the observed dynamics involves the interplay of
populations with distinct orbital characteristics. 
Although such a populations-based approach may presently be insufficient
to explain the velocity
dispersion gradient in a way envisioned by McConnachie et al.,
ultimately unraveling this interplay may be the best path to
a definitive assessment of the underlying mass distribution.  As pointed out by
Walker \& Pe\~narrubia (2011) (see also Agnello \& Evans 2012), that each subpopulation
is in equilibrium within the same gravitational potential can be used to measure directly the slope
of the density profile using simple mass estimators --- but {\it only} 
if the triaxial orientation of the satellite
can be determined (Kowalczyk et al. 2013).  The latter is a prospect, however, that is particularly promising
for the Sgr system ({\L}okas et al. 2010a; Kowalczyk et al. 2013).

\acknowledgements
We gratefully acknowledge support by National Science Foundation (NSF) grant AST11-09718
and by the
Polish National Science Centre under grant NN203580940 to EL{\L}.
 
Funding for SDSS-III has been provided by the Alfred P. Sloan Foundation, the Participating Institutions, the National Science Foundation, and the U.S. Department of Energy Office of Science. The SDSS-III web site is http://www.sdss3.org/.

SDSS-III is managed by the Astrophysical Research Consortium for the Participating Institutions of the SDSS-III Collaboration including the University of Arizona, the Brazilian Participation Group, Brookhaven National Laboratory, University of Cambridge, Carnegie Mellon University, University of Florida, the French Participation Group, the German Participation Group, Harvard University, the Instituto de Astrofisica de Canarias, the Michigan State/Notre Dame/JINA Participation Group, Johns Hopkins University, Lawrence Berkeley National Laboratory, Max Planck Institute for Astrophysics, Max Planck Institute for Extraterrestrial Physics, New Mexico State University, New York University, Ohio State University, Pennsylvania State University, University of Portsmouth, Princeton University, the Spanish Participation Group, University of Tokyo, University of Utah, Vanderbilt University, University of Virginia, University of Washington, and Yale University.

\end{document}